\documentstyle[twoside,epsfig,fleqn,espcrc2]{article}
\def\beq{\begin{equation}}
\def\eeq{\end{equation}}
\def\eps{\epsilon}
\title{Instanton Distribution in Quenched and Full QCD}

\author{%
R.C.~Brower,\address{Department of Physics, Boston University,
    Boston, MA 02215, USA} 
T.L.~Ivanenko,\address{Center for Theoretical Physics,
Massachusetts Institute  of Technology, Cambridge, MA 02139, USA}
J.W.~Negele,\hbox{$^{\rm\alph{address}}$}
K.N.~Orginos\address{Department of Physics, Brown University,
    Providence, RI 02912, USA}%
\thanks{Based on the talk presented by T.L.I. 
Work supported by U.S. Department of
Energy (D.O.E.) under cooperative research agreement
DE-FC02-94ER40818.}}

\begin{document}

\begin{abstract}
In order to optimize cooling as a technique to study the instanton
content of the QCD vacuum, we have studied the effects of alternative
algorithms, improved actions and boundary conditions on 
the evolution of single instantons and instanton anti-instanton
pairs. Using these results, we have 
extracted and compared the instanton content of quenched and full QCD.
\end{abstract}

\maketitle

\section{Introduction\label{sec::intro}}

Because of the significant role that instantons play in light hadron
structure \cite{Shuryak,Chu} and the intrinsic importance of
understanding topological excitations in QCD, it is of fundamental
interest to understand the instanton content of the QCD vacuum and, in
particular the effect of dynamical quarks on it. Although cooling is a
powerful technique to preferentially filter out non-topological
excitations relative to instantons, it is limited by the removal of
instantons due to lattice artifacts and instanton anti-instanton annihilation.
We address this problem and compare quenched and 
unquenched results.

\section{Relaxation Algorithm\label{sec::relax}}

In order to extract the instanton content of the lattice
configurations efficiently on a parallel computer, we used a variant
of the cooling method by discretizing  the relaxation equation,
\beq
U^\dagger\frac{\mbox{d}U}{\mbox{d}\tau} = -\frac{\delta S}{\delta U} ,
\label{eq::relax}
\eeq
introducing a small step size parameter $\epsilon$ in the
``relaxation time'' $\tau$ and updating 
simultaneously {\it all} links on the lattice.
For large values of $\eps$ this algorithm is unstable but in the
limit $\eps\rightarrow0$ it converges to the solution of the
relaxation equation(\ref{eq::relax}). We have found that for SU(3)
gauge field the value $\eps=0.025$ gives fast and stable relaxation
and we have used this value in our measurements. 
Comparing the evolution of the action for our relaxation method with
$\eps=0.025$ and the ``standard'' Cabbibo-Marinari cooling, we found that
one cooling step is approximately equivalent to 4 relaxation steps and
that the cooling histories are very similar.

\section{Identifying Instantons}

We used the following procedure to identify instantons in our
configurations. We consider all peaks in the action and topological
charge density as candidates for the instantons.  With an initial
value of the instanton position $x_0 $ (peak position) and instanton
size $\rho_0 =(48/S_0)^{1/4}$ (with the normalization
$S(x)\rightarrow a^4 F\tilde{F}$ in the continuum), we perform a least
squares fit to the action and topological charge densities of a
classical continuum instanton,
\beq
    S_0(x,x_0)=\frac{48\rho^4}{((x-x_0)^2+\rho^2)^4}.
    \label{eq::class-inst}
\eeq
as a function of $x_0,\rho$.  If the fitting process converges and the
final values for the instanton position $x_0$ and size $\rho$ are
within a pre-defined range close to the initial values, we record the
identified instanton. If one of the conditions above is not satisfied,
the candidate peak in the action or topological charge is discarded.

Full details of the algorithm will be published
elsewhere\cite{TBP}. Although there is some degree of arbitrariness in
selecting the parameters in the algorithm, we were able to identify
around 50\% of the peaks in the action distribution as instantons with
the remaining peaks corresponding to small instantons or other
excitations.

\section{Isolated Instantons\label{sec::1inst}}

Several effects can distort the instanton distribution during the
relaxation process.  The smallest instantons may disappear by falling
through the mesh. The larger instantons may shrink because of
interaction with periodic images or may disappear by
instanton anti-instanton pairs annihilate. We separated these effects
by studying the evolution of discretized semi-classical instanton
configurations.

\begin{figure}[htb]
\vspace*{7pt}
\epsfig{file=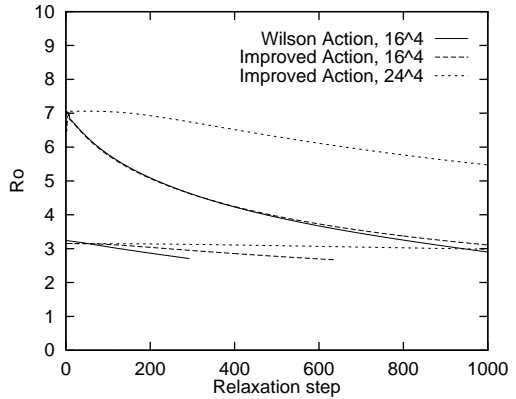,clip=}
\vspace*{-1.0cm}
\caption{Evolution of isolated instantons size $\rho=3.0$ and
  $\rho=7.0$ under relaxation for Wilson and  improved actions on
  lattice sizes $16^4$ and $24^4$.}
\label{fig::one-inst}
\vspace*{-0.6cm}
\end{figure}

First, we have investigated the shrinking of a single isolated
instanton for a variety of instanton sizes and two lattice sizes,
$16^4$ and $24^4$. 
We studied two actions, the Wilson action and a first order
improved action~\cite{imp-cooling},
\beq
S_{\mbox{imp}}=\frac{4}{3}\mbox{Tr}(1-W_{1\times1}) -
\frac{1}{48}\mbox{Tr}(1-W_{2\times2}).
\eeq
The first order term in the Wilson action expansion in $a^2/\rho^2$ is
negative and the relaxation process forces instantons to shrink. The
coefficients in the improved action were chosen so that the first
order term is zero and the second order term is positive. In
Fig.~\ref{fig::one-inst}, we show the evolution of a single isolated
instantons with sizes $\rho=3.0$ and $\rho=7.0$. Notice that during
relaxation the trajectories terminate for the smaller instantons as
they ``fall through the lattice'' and that the small instantons are
much more stable with the improved action \cite{imp-cooling}. Adding
more terms to the action could further stabilize instantons
\cite{imp-actions} but the first-order improved action was simple and
already sufficient for our studies. For large instantons the evolution
does not depend appreciably on the choice of the action but the
boundary effects are substantial.  Since we are ultimately interested
on the order of 100 relaxation steps at most, we conclude that with
the improved action and significantly larger lattices, the artificial
loss of single instantons in cooling is negligible.

\begin{figure}[htb]
\vspace*{7pt}
\epsfig{file=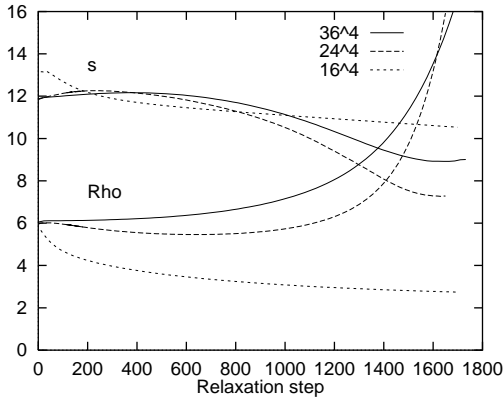,clip=}
\vspace*{-1.4cm}
\caption{Evolution of I-A pair under relaxation on the lattices of
  different size. Upper curves gives the I-A separation $s$ and the
  lower the sizes~$\rho$. }
\label{fig::two-inst}
\vspace*{-0.3cm}
\end{figure}

Next, we studied the evolution of an instanton anti-instanton (I-A)
pair.  Figure~\ref{fig::two-inst} shows the history in a typical case:
instantons of initial sizes $\rho_I=\rho_A=6.0 a$ with the separation
between centers $s=12.0 a$. On the small lattice $16^4$, the boundary
effects are much stronger than the interaction between objects and
they shrink in place before they have a chance to interact. On the
larger lattices the instanton and anti-instanton attract each other and
annihilate. The finite volume effects have stabilized on the largest
lattices.  We have also checked that the discretization effects are
negligible by comparing the evolution of the two appropriately
rescaled configurations. (In Fig~\ref{fig::two-inst} the
``data'' for the $36^4$~lattice are actually obtained by a re-scaling
from a $24^4$~lattice.)

The ``interaction'' time during which the instanton move toward each
other is much less then the ``shrinking'' time and the individual
instantons in a pair do not shrink significantly. In the case of a
real dynamical configurations, we expect the instantons to interact
more strongly and hence we believe the conventional use of periodic
boundary conditions will not introduce serious finite volume
distortions.

\section{QCD Results\label{sec::MC}}

We have studied 21 full QCD configurations for $\beta=5.5$, 2 flavors
of Wilson fermions and $\kappa=0.160$. As measured by LANL
group~\cite{LANL2}, this choice of parameters corresponds to the
lattice spacing $a(f_\pi)=0.11 \mbox{fm}$ and pion mass $m_\pi=360
\mbox{MeV}$. Configurations were separated by 50 HMC trajectories with
average length 50 steps of size $\epsilon=0.01$.  Those configurations
have the same value of $\kappa_c=0.16145$ as quenched QCD for
$\beta=5.858$ (See Ref.~\cite{LANL2}).  To compare quenched and full
QCD in roughly the same physical region, we have generated 23 quenched
configurations for $\beta=5.85$ separated by 500 heat-bath iterations.
We checked also that the Creutz ratios in both cases are comparable in
the range of 3-5 lattice units.

\begin{figure}[htb]
\vspace*{-0.5cm}
\epsfig{file=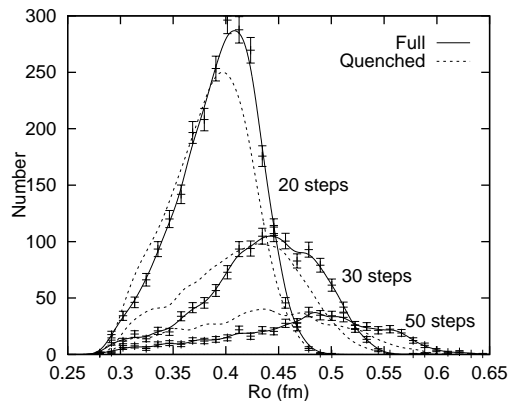,clip=}
\vspace*{-1.3cm}
\caption{The instanton distribution in quenched and full QCD. The smooth
  curve is the measured distribution convoluted with a Gaussian curve
  of width 0.01fm. Error bars are estimated by the  jackknife method.}
\label{fig::distrib-both}
\end{figure}

Figure~\ref{fig::distrib-both} shows the distribution of instantons in
sizes after 20, 30 and 50 relaxation steps. Since our method for
identifying instantons is not reliable before 20 steps, we can not
determine the early evolutions of the distribution, but after 20 the
distribution certainly changes dramatically with relaxation.  Given
that the number of instantons are $O(200)$ or larger in a  $16^4$ box,
the average distance between instantons is about 4 lattice units,
comparable to the average instanton size. Under these conditions the
interaction between instantons is very strong and consequently the
effects of I-A pair annihilation are much stronger than boundary
effects.

Since the erosion of the instantons distribution with cooling should
be the same for the quenched and unquenched samples, we believe the
fact that the distributions shown in Fig.\ref{fig::distrib-both} are
essentially identical within errors provides  strong evidence that the
physical instanton distributions are very similar in  quenched and full
QCD, at least at this sea  quark mass, which is of the order of $m_s$.

\end{document}